\documentclass[12pt]{article}
\usepackage{epsfig,latexsym}
\date{\today}
\def\bm{\boldmath}
\def\be{\begin{equation}}
\def\ee{\end{equation}}
\def\bear{\be\begin{array}}
\def\bea{\begin{eqnarray}}
\def\eea{\end{eqnarray}}

\def\dst{\displaystyle}
\def\lsi{\raise0.3ex\hbox{$<$\kern-0.75em\raise-1.1ex\hbox{$\sim$}}}
\def\gsi{\raise0.3ex\hbox{$>$\kern-0.75em\raise-1.1ex\hbox{$\sim$}}}
\def\lsim{\mathop{\lsi}}

 \def\Pom{{\bf I\!P}}
\def\od{{\cal O}}

\newcommand{\fr}[2]{\frac{{\dst #1}}{{\dst #2}}}
\newcommand{\fn}[1]{\footnote{{ #1}}}

\newsavebox{\fmbox}
    
\title {\bf \bm  Possibility of the odderon discovery via
observation  of charge asymmetry  in the diffractive $\pi^+\pi^-$
production  at HERA}

 \author{I.F.~Ginzburg$^{1}$\thanks{E-mail:
 ginzburg@math.nsc.ru},
 I.P.~Ivanov$^{1,2}$\thanks{E-mail: i.ivanov@fz-juelich.de},
 N.N.~Nikolaev$^{2,3}$\thanks{E-mail: n.nikolaev@fz-juelich.de}\\
 \makebox[8cm][l]{\normalsize
 $^1$  Institute of Mathematics, Novosibirsk, Russia}\\
 \makebox[8cm][l]{\normalsize
 $^2$  IKP, Forschungszentrum J\"ulich, Germany}\\
 \makebox[8cm][l]{\normalsize
 $^3$  L.D.~Landau ITP, Moscow, Russia}\\
 }

\begin{document}

\maketitle

 \vspace{-9cm}
 \makebox[\textwidth][r]{\large\bf FZJ-IKP(Th)-2002/15}
 \vspace{7.5cm}

\abstract{ The interference between the Pomeron and possible
odderon mechanisms of diffractive $\pi^+\pi^-$ photoproduction
results in charge asymmetry of the produced pions. The
observation of charge asymmetry of pions at moderate
$M_{\pi^+\pi^-}$ will be an undoubted signal of odderon existence.
To make numeral estimates more definite, we limit ourselves by the
region $M_{\pi^+\pi^-} = 1.1 \div 1.5$ GeV, where in the odderon
mechanism of dipion production, the production via single
$f_2(1270)$ resonance is expected to be dominant. 
We find a very statistically significant
effect of the odderon induced charge asymmetry even with very
modest estimates for the $f_2$ photoproduction cross section
(without referring to any particular model of the odderon).

\section{Introduction}

Pomeranchuk's conclusion that the particle and antiparticle cross
section differences vanish at asymptotic energies as compared to
the cross sections themselves is well known \cite{Chuck}. As early
as in 1970 there were debates \cite{Gershtein} that certain
particle-antiparticle cross section differences might not vanish
with energy growth, and properties of amplitudes not satisfying
the conditions of the Pomeranchuk theorem have been investigated
to much detail by Gribov et al. \cite{Gribov}. Later on, the term
{\sl odderon}, $\od$\  has been coined \cite{Luk} for the
singularity with $C=-1$ and the intercept $\alpha_\od \approx 1$.
Because the particle-antiparticle cross section difference
$\sigma^{+}- \sigma^{-}$ should not exceed the sum $\sigma^{+}+
\sigma^{-}$, the intercept of the odderon is expected to be not
higher than that of the Pomeron, $\alpha_\od \le \alpha_\Pom$.

\subsection{The theoretical and experimental status of the
odderon}

The odderon is an integral feature of the QCD motivated picture of
high energy scattering, and its experimental discovery is crucial
for the QCD description of strong interactions. Within
perturbative QCD (pQCD), the Pomeron  exchange is naturally modeled by
the color-singlet two-gluon exchange in the $t$-channel
\cite{BFKL} and at the same very level the odderon is modeled by
the $d$-coupled, color-singlet, three-gluon exchange in the
$t$-channel \cite{Bart,Landshoff}, which suggests that both
Pomeron  and odderon intercepts are close to the gluon spin,
$\alpha_{{\od}, \Pom} \sim J_{g}=1$. The experimental data and
the BFKL calculations show that the Pomeron  intercept $\alpha_\Pom(0)
>1$ (for the recent work and references see \cite{BFKLP}). The
theoretical estimates for $\alpha_{\od}$ are not yet conclusive,
the published results for the $\alpha_s= const$ approximation
grow gradually with time: $\alpha_{\od}(0)=0.94\to 0.96\to 1\to
?$, see e.g. \cite{Lipatov} and references therein. In our
discussions we will keep in mind that both $\alpha_\Pom$ and
$\alpha_\od$ are close to 1, and $\alpha_\Pom-\alpha_\od$ is
small.

If the Pomeron  and the odderon are assumed to be Regge poles,
their contributions to the scattering amplitude $AB\to CD$ have
the standard factorized form
 \bear{c}
A_{\cal R}=\zeta({\cal R})e^{i\pi \alpha_{\cal R}/2}\cdot G(A{\cal
R}C) \cdot s^{\alpha_{\cal R}}\cdot G(B{\cal R}D)\,,\\[1mm] \mbox{
with }\;{\cal R}=\Pom,\,\od\,,\quad\zeta(\Pom)=1\,,
\quad\zeta(\od)=i\,.
\end{array}\label{basform}
 \ee
Here the factors $G(A{\cal R}C)$ and $G(B{\cal R}D)$ describe the
couplings $A{\cal R} C$ and $B{\cal R}D$ respectively. They depend
on the particle helicities $\lambda_i$, and at small $t$ one must
have
 \be
G(A{\cal R}C)\propto |t|^{|\lambda_A-\lambda_C|/2}\,,\quad
G(B{\cal R}D)\propto|t|^{|\lambda_B-\lambda_D|/2}
\,.\label{bashel}
 \ee
Since the intercepts $\alpha_\Pom$ and $\alpha_\od$ are close to
1, the Pomeron exchange amplitude is predominantly imaginary,
while the odderon exchange amplitude is predominantly real.

$\diamondsuit$ Although in the pQCD framework the Pomeron and
odderon are of a similar status, the experimental quest for the
odderon exchange has proven to be a challenging task. The
estimates for the cross section difference $\sigma_{pp}-
\sigma_{p\bar{p}}$ turned out to be well below the experimental
uncertainties. The related studies of $Kp$ scattering are possible
only at fixed target in the limited range of $s \lsim 10^3$
GeV$^2$ and with relatively low statistical accuracy. The
diffractive photoproduction of $C=+1$ (pseudo) scalar and tensor
mesons $M$, $\gamma p \to p' M$ (with $p'$ either proton or its
low-mass excitation) suggested a decade ago \cite{givmesons},
seems to be a better signature for the odderon exchange. Indeed,
the $C=+1$ mesons are excited from $C=-1$ initial photons only via
the $C=-1$ exchange in the $t$-channel. For sufficiently large energies
the $\rho$ and $\omega$ exchange contributions die out (see
Appendix for details), and such processes will be dominated by the
odderon exchange and --- for the expected $\alpha_\od\sim 1$ ---
have the cross section, which is approximately flat vs. energy.
The systematic search for such reactions with flat cross sections
is in its formative stage, and they have not yet been observed
experimentally. These results at least suggest that the odderon is
weakly coupled to the proton as has been argued earlier (see e.g.
\cite{Zakharov,IFGod}).

$\nabla$ Available theoretical calculations for the soft
odderon amplitudes at small $t$ \cite{Zakharov,Zhitnitsky,Berger}
are based on variety of nonperturbative 3-gluon exchange and
nonrelativistic quarks models for mesons and nucleons. Even in
hard electroproduction of pion pairs \cite{hardpi} or
photoproduction of open charm \cite{brodsky}, one cannot eliminate
the sensitivity of the odderon amplitude to the soft quark models
of the proton, so that, calculations of \cite{brodsky,hardpi} can
be regarded only as crude estimates with large uncertainty. For
example, the ratio of absolute values of the forward odderon and Pomeron
exchange $NN$ amplitudes varies from $0$ to $0.04$
depending on the clustering of quarks in the
nucleon \cite{Zakharov}. Similar estimates for the exclusive
$\pi^0$ photoproduction lead to the cross sections varying from 10
to 200 nb \cite{Zhitnitsky,Berger}. According to
ref.~\cite{Zakharov}, in the reggeized 3-gluon exchange
quark--diquark model the diagonal $p{\od} p$ vertex disappears in
the unrealistic limit of the point-like scalar diquark (and in
this limit $p{\od} p'$ vertexes with proton excitations $p'$
become dominant), while the diagonal vertex become essential or
even dominant with the growth of the diquark size (unfortunately, this
very approximation of the point-like scalar diquark was used in
ref.~\cite{Berger}). This type of uncertainty makes all available
estimates for odderon rather dubious.

$\nabla$ Another important limitation of the above cited
calculations is that they were performed only in the Born
type approximation. In the reggeization program, which is technically
carried out by resummation of leading logarithms of energy from
loop corrections, the Born terms are just the starting point. It
remains just an assumption, although ---
since the odderon intercept is close to unity--- a plausible one, that the
Born result sets a reasonable scale for the reggeized physical
amplitude.

Still, such estimates must be taken with the grain of salt.
Indeed, the trademark of the reggeon amplitude is the
factorization (\ref{basform}), with factorized polarization
dependence of the form~(\ref{bashel}). The Born approximation
results usually contain both helicity factorized and helicity
non-factorized terms. After the resummation, the factorized terms
give rise to the Regge behavior, while the contribution from
non-factorized terms should vanish at $s\to \infty$. Consequently,
only the factorized components of the Born amplitude can be taken
for an estimate of the odderon exchange. This picture is supported
by direct calculations in all known cases, for a recent example
of the leading $\log s$ resummations see \cite{BalLip}. Such a
separation between factorized and non-factorized terms was not
performed in refs.~\cite{Berger}, as it can be clearly seen from
the results for the $\gamma p \to f_2p'$ photoproduction. Namely,
the leading term for the Born amplitude evaluated in \cite{Berger}
corresponds precisely to the non-factorizing amplitude with
correlated spin flips in the both vertices ($\lambda_M-
\lambda_\gamma =\lambda_p-\lambda_{p'}=1$), so that this amplitude
does not vanish at $t=0$, in contrast to (\ref{bashel}). According
to the above arguments, such a term must decrease with energy
after resummation and therefore must be removed from the discussed
result. Therefore, two essential conclusions of \cite{Berger}
cannot definitely  be related to the odderon contribution:\\
 {\it (i)} the values of the cross section estimated;\\
 {\it (ii)} the prediction of the nucleon excitation dominance for
the proton vertex.

$\diamondsuit$ Recently, the H1 collaboration reported first
results on the search for the odderon in diffractive
photoproduction of $C=+1$ mesons. The event selection was inspired
by results of \cite{Berger} and included only events with nucleon
excitations. No signal from the odderon was found, and the upper
limits on the cross sections were placed at $\sigma (\gamma p\to
\pi^0X<39$ nb \cite{H1pizero}, $\sigma(\gamma p\to f_2 X) < 16$ nb
and $\sigma(\gamma p\to a_2 X) < 96$ nb (both in \cite{H1f2}).
These bounds are below the correspondent predictions of
refs.~\cite{Berger}. In the light of the above discussion, this is
hardly surprising. In particular, we see no reason whatsoever to
perform any specific selection regarding the nucleon final state.

Below we do not cling to any specific model for the odderon except
for very natural assumptions about similarity of the odderon to
the other reggeons, and estimate observable effects only by
assuming that the odderon mechanism of $C$--even dipion production
is larger than the non--odderon mechanisms (see Appendix for
details). In this respect we use the standard Regge-pole model for
the Pomeron and odderon mediated amplitudes that describe the
diffractive production of a $\pi^+\pi^-$ pair in the $C$-odd and
$C$-even states respectively, and assume that these pion pairs are
produced via intermediate resonance states.

\subsection{Exploiting the charge asymmetry}

The chances of discovery of the elusive odderon are arguably
enhanced if instead of isolation of the pure odderon exchange
reactions one would look for the Pomeron--odderon interference,
which is linear in the small odderon amplitude (not quadratic as
contribution to cross section). This interference can be observed
as the {\it charge asymmetry} of diffractively produced particles.
The main idea can be formulated as follows. The initial photon has
definite $C$-parity, $C=-1$. Since the Pomeron ($\Pom$) has vacuum
quantum numbers, i.e., $C=+1$, the Pomeron--photon collision
produces $C$--odd system. To the contrary, the collision of photon
with $C=-1$ odderon ($\od$) produces $C$--even systems.
Consequently, it is useful to study the production of final system
that can be produced both by the Pomeron and the odderon
exchanges. The interference of the corresponding $C$-odd and
$C$-even amplitudes gives rise to a charge asymmetry in the
momentum distribution of produced particles. In the absence of the
other $C$-odd exchange mechanisms, it will be an unambiguous
signal of the odderon.

The search for the odderon via the charge asymmetry in the
photoproduction of $c\bar{c}$ pairs was proposed first in
ref.~\cite{brodsky}. The obvious disadvantage of this process is
the small diffractive cross cross section, further hampered by a
small efficiency of detection of charmed particles (see \cite{Ira}
and references therein). Note that the final estimates obtained in
this paper have large uncertainties due to above mentioned
uncertainties in the description of odderon--nucleon vertex.
Besides, under standard assumptions about the quark--hadron duality
for heavy quarks the asymmetry obtained disappears at
$\alpha_\od\to\alpha_\Pom$,
 \be
Re\left({\cal A}_\Pom^\dag {\cal A}_\od\right)\propto Re\left\{ i
\exp\left[{i\pi(\alpha_\Pom-\alpha_\od)\over 2} \right]\right\} =
\sin\left[{\pi(\alpha_\Pom-\alpha_\od)\over 2}\right]\,.
\label{eq:1.1}
 \ee
In sect.~4.1 we show that this conclusion becomes invalid due to
the final state strong interaction (FSI).

About two years ago we suggested\fn{ At different stages, the
preliminary results were published in ref.~\cite{GIN1} and have
been repeatedly reported during last two years \cite{review}} to
look for the charge asymmetries in the much more copious
diffractive photoproduction of $\pi^+\pi^-$ pairs at
$M_{\pi^+\pi^-} \lsim 1\div 1.5$ GeV. The advantages of this
process as compared to the $c\bar{c}$ production are\\ ({\it i})
the much higher basic cross sections and high detection efficiency
for pions and \\
({\it ii}) the final state interaction (FSI) is essential and the
dipion production amplitudes $F_\Pom$ and $F_\od$ acquire
additional phase shifts  as compared to (\ref{eq:1.1})  (for the
early discussion on FSI effects within the Regge formalism see
\cite{Anisovich}), which are predominantly controlled by the
prominent pion-pion resonances. Zooming in on the mass region
where the Breit-Wigner phase shifts are such as the small factor
(\ref{eq:1.1}) is eliminated, one can gain the charge asymmetry
that would persist even if $\alpha_\Pom= \alpha_\od$.

Recently, the idea that the odderon can be discovered via
observation of charge asymmetry of pions in diffractive
$\pi^+\pi^-$ was extended to hard electroproduction \cite{hardpi}.
Certainly, the cross sections calculated in that work  are
much lower than the photoproduction cross sections discussed here.
Besides, the numerical estimates of this paper are unsafe due to
the above mentioned uncertainty in the description of $p\od p'$
vertex. Last, the statement about the dominance of transverse
charge asymmetry is doubtful, since the value of photon virtuality
$Q^2\approx 3$ GeV$^2$ does not seem to be high enough for a
definite statement regarding the dominance of the leading twist
amplitudes. Note that with growth of the electron scattering
angle, the contribution of $Z$ exchange increases. The axial
component of $Z$ current can produce C--even dipions in fusion
with Pomeron as well. This effect can overshoot the small odderon
effect at large enough $Q^2$.

Our approach is similar in some respect to that used for the
description of charge asymmetry in the process $e^+e^-\to
e^+e^-\pi^+\pi^-$ \cite{CherSer,SerboAsym}, which is suitable for
the study of low energy phenomena and resonances in pion and kaon
physics.

The paper is organized as follows. In Sect.~2 we introduce
notation, define the forward-backward (FB) and the transverse (T)
asymmetries. In Sect.~3 we describe the Pomeron and odderon
helicity amplitudes for dipion production via intermediate
resonance state. The charge asymmetry due to Pomeron--odderon
interference is calculated in Sect.~4. In Sect.~5 we present
numerical estimates which appear very promising. Discussion of the
results obtained and the conclusions are presented in Sect.~6.
Last, in the Appendix we discuss non-odderon mechanisms of $C=+1$
meson production and find the lowest value of odderon mediated
cross section that --- provided natural cuts are applied ---
cannot be mimicked by non-odderon mechanisms.

\section{Kinematics}

We focus on the real photoproduction reaction $\gamma p\to
\pi^+\pi^- p'$ with energies lying in the HERA range. The pion
system with a small to moderate invariant mass $M$ is produced
with a large rapidity gap from the recoil proton $p'$. The initial
momenta of the photon and proton are $q$ and $P$ respectively,
$s=(q+P)^2$, initial photon polarization is $\vec{e}$. With $k_+$
and $k_-$ being the momenta of the charged pions, we consider
\be
r^\mu = k^\mu_+ - k^\mu_-\,,\quad \Delta^\mu = k^\mu_+ +
k^\mu_-\,, \quad M^2=\Delta^2\,.
 \ee
The discussed charge asymmetry must appear precisely as the
antisymmetry of the differential cross sections under replacement
$r^\mu \to -r^\mu$.

We perform calculations in the helicity basis where
$\lambda_\gamma$ and $\lambda_R$ are the helicities of photon and
produced dipion respectively, while $\lambda_p$ and $\lambda_{p'}$
are the helicities of incident and scattered proton respectively.
The final results are averaged over initial photon polarizations.

We define the $z$-axis as the $\gamma p$ collision axis.
Throughout the paper 2D transverse vectors will be marked with the
vector sign, while the 3-vectors will be given in bold. We direct
the $x$-axis along vector $\vec{\Delta }$ and define by $\psi$ the
azimuthal angle of the vector $\vec {\Delta}$, i.e., the
production plane, with respect to the fixed lab frame of
reference, in which the helicity states of the incoming photon are
defined. For instance, for the tagged photons in electroproduction
$e p \to e \pi^{+}\pi^{-}p'$, this frame of reference can be
related to the electron scattering plane. Then the polarization
vector of the initial photon with helicity $\lambda_\gamma=\pm 1$
can be written as $\vec e^{\lambda} = -{1 \over\sqrt{2}}\cdot
e^{i\lambda_\gamma \psi} (\lambda_\gamma, i)$. Hereafter we
neglect the pion mass compared to the mass of dipion $M$.

To define the appropriate independent charge-asymmetric
observables, let us first denote by $z_+$ and $z_-$ the standard
light cone variables for each charged pion, $z_\pm=
(\epsilon_\pm+p_{\pm z})/(2E_\gamma)=(k_\pm P)/(qP)$ (for the
considered diffractive type processes $z_++z_-=1$). Then we define
appropriate variables for the description of charge asymmetry
\be
\xi=\fr{z_+-z_-}{z_++z_-}\,,\quad v=
\fr{2(\vec{k}_+^2-\vec{k}_-^2-\xi\vec{\Delta}^2)}{M|\vec{\Delta}|}
\equiv \fr{2(\vec{\rho}\vec{\Delta})}{M|\vec{\Delta}|} \;\mbox{
with } \vec\rho =\vec{r}-\xi\vec{\Delta}\,  . \label{asyms}
 \ee
(Note that the transverse momentum of each charged  pion is split
in two parts as $\vec{k}_\pm=\pm (\vec{\rho}/2)+ z_\pm\vec{\Delta}
$. Here  $\vec{\rho}/2$ is relative transverse momentum of $\pi^+$ in
respect to the total transverse momentum of the dipion.) With such
definition, the variables $\xi$ and $v$ describe the
forward-backward (FB) and transverse (T) asymmetries of the
charged pions respectively. In terms of these variables the
discussed charge asymmetry is non-invariance of the differential
cross section under the transformation $\xi\to -\xi $ and $v\to -v
$. For example, positive transverse asymmetry means that the
number of events with $v>0$ exceeds that with $v<0$.

For the  pion pair production the above observables $\xi$ and $v$
are related simply to the polar and azimuthal angles $\theta$ and
$\phi$ of the 3-momentum $\bf{k}_+$ in the dipion rest frame $R$
 \be
\left.{\bf k}_+\right|_R = {1\over 2}  (\vec{\rho},M\xi) = {1\over
2}M(\sin\theta \cos\phi,\; \sin\theta \sin\phi,\; \cos\theta)\, .
\label{eq:2.3}
 \ee
In this frame for $\bf{k}_-$ we have the same equations with
$(\theta\,,\phi)\to (\pi-\theta\,,\pi+\phi)$. The charge
asymmetric variables (\ref{asyms}) are related to these angles as
\be
\xi=\cos\theta\,,\quad v= \sin\theta\cos\phi \,. \label{xiv}
 \ee

In the following we shall discuss the five-fold differential cross
section of the reaction $\gamma p \to \pi^{+}\pi^{-}p'$
 \be
2\pi\,\fr{d\sigma}{dM^{2}\, d\Delta^{2}\,  d\cos\, \theta\,
d\phi\,d\psi}\;\equiv\; 2\pi\, \fr{d\sigma }{dM^{2}\,
d\Delta^{2}\,d\xi\,dv\,d\psi}\cdot {1 \over 2}\sqrt{1-\xi^2-v^2}
\; .\label{phsp}
 \ee
Factor $1/2$ reflects degeneracy of $v$ in respect to
replacement $\phi \to 2\pi-\phi$. The integration measure in terms
of $\xi, v$ is
\be
\int d\Omega \equiv \int\limits_{-1}^1
d\cos\theta\int\limits_0^{2\pi}d\phi \quad\longrightarrow\quad
2\int\int\fr{d\xi\ dv} {\sqrt{1
-\xi^2-v^2}}\;\;\theta(1-\xi^2-v^2)\,.\label{measure} \ee

\section{The Pomeron and odderon helicity amplitudes}

In this section we consider the main features of the high energy
diffractive $\gamma p\to \pi^+\pi^- p'$ amplitudes from the
Pomeron and the odderon exchange.

\subsection{General properties}

$\bullet$ The properties of the {\bf Pomeron amplitude} are well
constrained by the experimental studies at HERA:

$\diamondsuit$ The main contribution to vertex $p \Pom p'$ comes
from the proton-elastic scattering, $p'=p$ (the admixture from
proton dissociation to excited states with masses $M' \lsim 2$ GeV
is well known to be below 25 \% \cite{AlberiGoggi,DoubleDiss}). To
a good approximation, the $s$--channel helicity conservation
(SCHC) holds for this vertex \cite{AlberiGoggi}.

$\diamondsuit$ The main contribution to the cross section is given
by amplitudes with production of two pions in the (C-odd)
$\rho$--meson state. At higher effective masses of dipion other
$\rho$ type resonances should also be accounted for. Besides, the
SCHC takes place at small $t$, i.e. the $\rho$--meson is produced
mainly with the same helicity as the helicity of the initial
photon.\vspace{2mm}

$\bullet$ There is no experimental information about the {\bf
odderon amplitude} and here one is bound to the theoretical
estimates.

$\diamondsuit$ In the reggeized 3-gluon exchange quark--diquark
model the spin properties of the odderon coupling to nucleons can
be summarized as follows \cite{Zakharov}:\\ ({\it i}) odderon
exchange satisfies SCHC and the spin-flip $p{\od} p$ amplitude
vanishes in the $SU(6)$ nonrelativistic quark model,\\ ({\it ii})
the relative strength of the spin-flip $p{\od} p$ vertex depends
on the size of the scalar diquark clustering, it becomes the
dominant one in the unrealistic limit of the point-like scalar
diquark (unfortunately, this very approximation of the point-like
scalar diquark has been used in ref.~\cite{Berger} ).

Therefore, we assume that the properties of $p \od p'$ vertex are
roughly similar to those for Pomeron, i.e, the SCHC elastic
transition with $p'=p$ is not suppressed.

$\diamondsuit$ The vertex $\gamma\od \pi^+\pi^-$ is of main
interest to us. We assume that --- as it is customary for other
phenomena at $M\lsim 1.5$ GeV --- the pion pairs are produced
mainly via resonance states ($C$--even $f_0$ and $f_2$ mesons in
our case). Below we discuss how shape of the $C$-even $\pi^+\pi^-$
spectrum helps us separate out the odderon signal, and for
numerical estimates we take the $f_2(1270)$ meson. Since the
helicity structure of $\gamma\od\to f_2\to \pi^+\pi^-$ vertex is
not known, we will present results based on several limiting
cases.

\subsection{Detailed description}

The amplitudes of dipion production via production of resonances
$R$ of spin $J$ and helicity $\lambda_R$ (with SCHC in the proton
vertex) can be conveniently written in a factorized form:
 \be
{\cal A}_\Pom= A^{\lambda_R\lambda_{\gamma}}_\Pom \cdot
D_J(M^2)\cdot{\cal E}^{J,\lambda_R}_{\lambda_{\gamma}}\,,\;\;
{\cal A}_\od= iA^{\lambda_R\lambda_{\gamma}}_\od \cdot
D_J(M^2)\cdot{\cal E}^{J,\lambda_R}_{\lambda_{\gamma}}\,.
\label{basampdef}
 \ee
The additional factor $i$ for odderon amplitude is related to the
opposite signature of the odderon as compared with the Pomeron.

In these equations the first factor describes the Regge amplitude
of production of the resonance $R$ with helicity $\lambda_R$ (the
energy dependence is included in the quantity $\sigma_R$)
 \be
A^{\lambda_R \lambda_{\gamma}}_{\mbox{{\tiny Regge}}} =
g^{\lambda_R}\sqrt{\sigma_R B_R}e^{i{\pi\over 2}\alpha_{{\cal
R}}}e^{-{1\over 2}B_R\vec{\Delta}^2}
\fr{(\sqrt{B_R}|\vec{\Delta}|)^{|\lambda_\gamma-\lambda_R|}}
{\sqrt{|\lambda_\gamma-\lambda_R|!}}\qquad (\mbox{Regge}
=\Pom,\,\od) \,.\label{resampl}
 \ee
Here $(g^{\lambda_R})^2$ is the fraction of total cross section of
the production of resonance $R$ with helicity $\lambda_R$ by the
photon with helicity $+1$. Since only helicity difference is
essential, this very quantity describes the transition of photon
with helicity $-1$ to dipion with helicity $-\lambda_R$ as well.
Since SCHC approximately holds for the Pomeron, we have
$g^1_\rho\approx 1$ and $|g^0_\rho|\ll g^1_\rho$. The $t$
dependence of this amplitude comes mostly from the vertex factors
(\ref{bashel}). It can be parameterized by
$\exp(-B_R\vec{\Delta}^2/2)$ modulo to helicity-flip factors
$|t|^{\Delta\lambda/2}$. For the sake of simplicity, we suppress
the $M$ dependence of the diffraction slopes $B_\rho$ and $B_{f}$
and the $t$-dependence of trajectories $\alpha_{\cal R}$.

The second factor in eq.~(\ref{basampdef}) describes the
$\pi^+\pi^-$ invariant mass dependence of the production amplitude
for the dipion state with angular momentum (spin) $J$ and directly
reflects the shape of the corresponding $C$-even or $C$-odd
$\pi^+\pi^-$ spectrum. In a simple case of single-resonance
dominance, it can be reasonably well approximated by the standard
Breit-Wigner factor for this resonance together with its coupling
to pions. Thus, for the ($C$-odd) $P$-wave $\pi^+\pi^-$ spectrum,
where the dominance of the $\rho$ meson is established, and for
the ($C$-even) $D$-wave $\pi^+\pi^-$ spectrum, where the
$f_2(1270)$ meson is expected to be the dominating feature, near
the resonance peak
 \be
D_J(M^2) = {\sqrt{m_R\Gamma_R Br(R\to\pi^+\pi^-)/\pi} \over M^2
-m_R^2 + im_R\Gamma_R}\,.\label{BW}
 \ee
For numerical estimates we use this form even away from the
resonance peaks, $|M^2-m_R^2|> M_R\Gamma_R$. The shape of the
$S$-wave $\pi^+\pi^-$ spectrum is expected to be more complicated.

Finally, the decay factor ${\cal
E}^{J,\lambda_{R}}_{\lambda_{\gamma}}$ describes the angular part
of the helicity amplitude. Because the pions are spinless, it
takes a particularly simple form,
\be
{\cal E}^{J, \lambda_{R}}_{\lambda_{\gamma}} =
Y_{J\lambda_R}(\theta,\phi)e^{i\lambda_{\gamma}\psi}\,,\label{anghel}
 \ee
where $J=1,2$ for the $\rho_0$ and $f_2$, respectively,
$Y_{lm}(\theta,\phi)$ is the standard angular momentum wave
function normalized as $ \int |{\cal E}^{J,
\lambda_{R}}_{\lambda_{\gamma}}|^2\; d\Omega\, d\psi/(2\pi) = 1$.

\section{Charge asymmetry}

\subsection{Invariant mass dependence}

The charge asymmetry effect is given by the interference of
suitable Pomeron and odderon amplitudes integrated over the
redundant phase space variables
 \be
d\sigma_{asym} = \sum\limits_{\lambda_R,\lambda_\rho}
2Re\left({\cal A}^{\lambda_\rho\dagger}_\Pom {\cal A}
^{\lambda_R}_O\right)d\Gamma\,. \label{dasym}
 \ee

The pattern of $M$-dependence of the  $\Pom-\od$ interference is
controlled by the $D_J(M)$ factors in the above amplitudes in the
form of {\em overlap functions}
 \be
{\cal I}_{12}(M^2) = Re\left[D_1(i D_2)^\dagger
 e^{i\delta_{\Pom O}}\right]\,,\quad
{\cal I}_{10}(M^2) = Re\left[D_1(i D_0)^\dagger
 e^{i\delta_{\Pom O}}\right]\,.
 \ee
The exact form of these overlap functions is given by the precise
form of different $D_J$ functions. Below, we will consider mainly
the region $M>1100$ MeV, where there are no $S$-wave dipion
resonances and, consequently, only the overlap function ${\cal
I}_{12}$ is of interest (see Figure 2 and related discussion). For
estimates here we use Breit--Wigner form (\ref{BW}) of $D_1$,
which is given by $\rho$--meson contribution from Pomeron, and
$D_2$, which is given by $f_2$--meson contribution from odderon
 \bear{c}
{\cal I}_{12}(M^2)
= Im\left(\fr{e^{i\delta_{\Pom O}}\sqrt{m_\rho
m_f\Gamma_\rho\Gamma_f
Br(f_2\to\pi^+\pi^-)Br(\rho\to\pi^+\pi^-)}}
{\pi(M^2-m^2_\rho+ i m_\rho\Gamma_\rho) (M^2-m_f^2-i m_f
\Gamma_f)}\right)\\[4mm]
 \mbox{ with } \delta_{\Pom O} = {\pi \over 2} (\alpha_{\Pom}
-\alpha_\od)\,.
\end{array}\label{overlap}
 \ee
In our analysis we neglect the $M$ dependence of diffraction
parameters, such as $B_\rho$, $B_f$, $g_R^{\lambda_R}$, etc. In
this approximation, the above overlap function is the only factor
that contains $M$ and it is universal for all asymmetries and $t$
dependencies.

\begin{figure}[!htb]
   \centering
   \epsfig{file=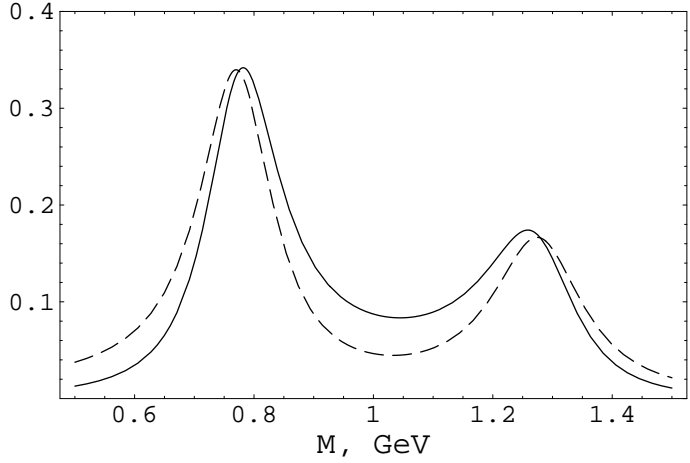,width=10cm}
   \caption{The $\rho-f_2$ overlap function ${\cal I}_{12}(M^2)$ calculated for
$\alpha_\Pom -\alpha_O = 0$ (solid line) and
$\alpha_\Pom -\alpha_O = 0.2$ (dashed line).}
   \label{fig1}
\end{figure}

Because the difference between Pomeron and odderon intercepts
$\delta_{\Pom O} $ is small, the overlap function is large only
when the phase shift between the two Breit-Wigner factors is close
to $\pi/2$. This will be the case in the vicinity of the resonance
peaks, where for the one resonance the $D_{J_1}$ is almost real
while for the other one the $D_{J_2}$ is almost imaginary, which
compensates the additional factor $i$ in the odderon amplitude.
The mass-dependence of the charge asymmetric overlap function is
shown in Fig.~\ref{fig1}. It exhibits only weak sensitivity to the
poorly known $\alpha_\Pom-\alpha_O$.
It is precisely the strong FSI that lifts
the suppression (\ref{eq:1.1}), which would come into play if FSI
were neglected.

\subsection{Final expressions}

We consider the differential cross sections averaged over initial
photon spin states and integrated over $\psi$, which is relevant
to $ep$ experiments with untagged scattered electrons. The
integration over $\psi$ leaves in the result only terms with
identical $\lambda_\gamma$. Besides, it is well known that for the
real photons the other factors depend only on
$|\lambda_\gamma-\lambda_R|$, not on the value of helicity itself.
Therefore, the observed interference effects will be proportional
to sums over opposite initial photon helicities with simultaneous
change of the sign of the produced resonance helicities, and will
have the azimuthal and polar angle dependence of the form $ {\cal
E}^{*J_\rho,\lambda_{\rho}}_{\lambda_{\gamma}}{\cal
E}^{J_R,\lambda_{R}}_{\lambda_{\gamma}}+{\cal
E}^{*J_\rho,-\lambda_{\rho}}_{-\lambda_{\gamma}}{\cal
E}^{J_R,-\lambda_{R}}_{-\lambda_{\gamma}} $. At odd $J_\rho-J_R$,
this changes sign under replacement $k_\pm\to k_\mp$ (
$\theta\to\pi-\theta$, $\phi \to\pi+\phi$), i.e. exhibits charge
asymmetry, either forward-backward or transverse.

Using the well known azimuthal dependence of spherical harmonics,
we have
 $$
 {\cal E}^{*J_\rho,\lambda_{\rho}}_{\lambda_{\gamma}}{\cal
E}^{J_R,\lambda_{R}}_{\lambda_{\gamma}}+{\cal
E}^{*J_\rho,-\lambda_{\rho}}_{-\lambda_{\gamma}}{\cal
E}^{J_R,-\lambda_{R}}_{-\lambda_{\gamma}}\propto
\cos[(\lambda_\rho-\lambda_R)\phi]\,.
 $$
This dependence give us the key to the {\em type of charge
asymmetry} that takes place for different helicities of the $\rho$
and $C$-even resonance $R$. The terms with odd $\lambda_\rho-
\lambda_R$ change sign under $\phi \to\pi+\phi$, {\em i.e.} under
$v\to -v$. They are responsible for the T asymmetry. The terms
with even $\lambda_\rho-\lambda_R$ remain invariant under $\phi
\to\pi+\phi$. Therefore, they must change sign under $\theta\to
\pi-\theta$, i.e. they are responsible for the FB asymmetry. In
other words, the interference of amplitudes with even
$|\lambda_\rho - \lambda_R|$ (0 or 2 in our examples) generates
the FB asymmetry, $\propto\xi P(\xi^2,v^2)$, whereas the
interference of amplitudes with odd $|\lambda_\rho - \lambda_R|$
(1 or 3) generates the T asymmetry, $\propto v P_1(\xi^2,v^2)$.

Starting from now, we will give expressions for the case of $J=1$
and $J=2$ interference ($\rho$--$f_2$) only. Similar formulas for
$J=1$ and $J=0$ interference (which are essential for the analysis
below 1.1 GeV) can be immediately obtained in the same manner.

Neglecting contributions with higher helicity flip,
$|\lambda_R-\lambda_\gamma|>1$, we obtain the C--odd interference
cross section of the form
 \bear{c}
\fr{d\sigma^{interf}}{dM^{2}\,
d\Delta^{2}\,d\xi\,dv}=\fr{3\sqrt{5}}{2\pi\sqrt{1-\xi^2-v^2}}
{\cal I}_{12}(M^2) \sqrt{\sigma_\rho \sigma_f B_\rho B_f}
\exp\left(-\fr{B_\rho + B_f}{2}|t|\right)\otimes T\\[4mm]
 T= g_\rho^1g_f^1 (1-\xi^2)\xi+vg_\rho^1\left[\fr{1}{2}g_f^2(1-\xi^2)+
\fr{1}{\sqrt{6}} g_f^0(3\xi^2-1)\right]\sqrt{B_f|t|}+\\[4mm]
 g_\rho^0g_f^1\sqrt{2}v\xi^2\sqrt{B_\rho|t|}+
 \xi g_\rho^0\left[\fr{1}{\sqrt{2}}g_f^2(2v^2+\xi^2-1)+
 \fr{1}{\sqrt{3}}g_f^0(3\xi^2-1)\right]\sqrt{B_fB_\rho}|t|\,.
 \end{array}\label{mainasym}
 \ee

$\bullet$ {\bf The forward--backward asymmetry} is obtained from
here by integration over $v$ (relative transverse motion of
pions):
 \bear{c}
\fr{d\sigma_{FB}}{dM^{2}\,
d\Delta^{2}\,d\xi}=\fr{3\sqrt{5}}{2} {\cal I}_{12}(M^2)
\sqrt{\sigma_\rho \sigma_f B_\rho B_f}\exp\left(-\fr{B_\rho +
B_f}{ 2}|t|\right)\otimes \xi\; T_\xi\,,\\[4mm]
 T_\xi=g_\rho^1g_f^1(1-\xi^2)+
\fr{1}{\sqrt{3}}g_\rho^0g_f^0(3\xi^2-1)\sqrt{B_fB_\rho}|t|\,.
\end{array}\label{xiasym}
\ee

If the SCHC holds for the odderon, then the principal effect would
be the FB asymmetry dominated by the first term in this equation.
If the mechanism of the $f_2$ production significantly violates
SCHC, then the first term is dominant only at small $t$. With the
growth of $|t|$, the terms with helicity flip for both the Pomeron
and odderon become essential, and generally, not small. Note that
upon the azimuthal integration the contribution from production of
$f_2$ in the state with helicity 2 vanishes because $\int
\cos2\phi d\phi=0$. Of course, one can isolate this contribution
to the FB asymmetry looking at the differential dependence on the
azimuthal variable $v$  or with integration over some region of
$v$, e.g. $v^2>v_0^2$.

$\bullet$ {\bf The transverse asymmetry} is obtained from
(\ref{mainasym}) by integration over $\xi$ (relative longitudinal
motion of pions) at fixed $v$:
 \bear{c}
\fr{d\sigma_T}{dM^{2}\, d\Delta^{2}\,dv}=\fr{3\sqrt{5}}{4} {\cal
I}(M^2) \sqrt{\sigma_\rho \sigma_f B_\rho
B_f|t|}\exp\left(-\fr{B_\rho + B_f}{2}|t|\right)\otimes v\;T_v
\,,\\[4mm]
 T_v=g_\rho^1 g_f^2 \sqrt{B_f}\;{1+v^2 \over 2}
 +g_\rho^1 g_f^0\sqrt{B_f} \sqrt{{2\over 3}}\;{1-3v^2 \over 2} +
 g_\rho^0 g_f^1\sqrt{B_\rho}\sqrt{2}(1-v^2)\,.
 \end{array}\label{vasym}
 \ee
Due to its kinematical $t$-dependence, the transverse asymmetry becomes
naturally small at small $t$ while the background is high here.
Therefore, imposing cuts from below in $|t|$ might improve the
signal to background ratio.

The transverse asymmetry is the dominant one in the case of SCHNC
for odderon, in particular, if the $f_2$ meson is produced mainly
in the state with maximal helicity $\lambda_f =\pm 2$. Evidently,
the similar transverse asymmetry is generated always when either
the SCHC odderon exchange interferes with the SCHNC helicity-flip
Pomeron exchange or vice versa. Certainly, one cannot exclude
accidental compensations among coefficients of these amplitudes.

The above analysis is similar to the well known partial wave
analysis (PWA). The eq.~(\ref{mainasym}) can be written in terms
of angular variables $\theta$, $\phi$, and above description can
be continued with PWA to a more detailed analysis of helicity
structure of the odderon amplitude (with known helicity structure
of the Pomeron amplitude) neglecting terms with large helicity
flips. To find all helicity flip amplitudes, one should consider
additionally dependence on initial azimuthal angle $\psi$, which
is measurable in the ep experiments (as it was done for the charge
symmetric contributions in ref.~\cite{Wolf}). Note that the
analysis of charge asymmetry in terms of asymmetry in $\xi$ and
$v$ is suitable also to multipion final states where PWA is very
complicated.

\section{Numerical estimates}

The main background to the charge asymmetry is given by the
Pomeron--photon (Pomeron--Primakoff) interference which is
predominantly transverse one since Primakoff mechanism produces
$f_2$ only in the states with helicity 2 and 0. To suppress this
background we suggest impose different cuts for the FB and T
asymmetries:
 \be
 |t_{FB}|=\vec{\Delta}^2\ge 0.1B_\rho^{-1}\approx 0.01\,\mbox{
 GeV}^2\,,\qquad
|t_T|=\vec{\Delta}^2\ge B_\rho^{-1}\approx 0.1\,\mbox{ GeV}^2\,.
\label{perplim}
 \ee

Below we use parameters of resonances from ref.~\cite{PDG} and
well known quantities for the $\rho$ meson photo--production,
$\sigma_\rho\approx 12\,\mu$b (for the diagonal in proton case,
$p'=p$), $B_\rho\approx 10$ GeV$^{-2}$, $g_\rho^1\approx 1$,
$g_\rho^0\approx 0.10$  (see refs.~\cite{H1spin,ZEUSspin} and
\cite{IgorSCHNC} for the data and their analysis). For the odderon
contribution we have no data. The estimates in Appendix show that
at HERA the odderon contribution would definitely dominate over
the other mechanisms if $\sigma_f\geq 1$ nb. Therefore, in order
to be able to make as strong conclusions as possible, we will take
the value $\sigma_f=1$ nb for the numerical estimates. Note that
this number is more than one order of magnitude smaller than the
currently discussed upper bound on the $\gamma p \to f_2 p$ cross
section. We hope that the real cross section is higher than our
very cautious estimate. We also take the slope parameter for the
$f_2$--meson $B_f=B_\rho$. We checked that the sensitivity of the
results to reasonable variations of $B_f/B_\rho$ is weak.

Obviously, the charge asymmetric contribution vanishes upon the
angular integration. We quantify the charge asymmetry by
 \be
\Delta\sigma_{FB} = \int d\sigma(\xi>0) - \int d\sigma(\xi
<0)\,,\qquad \Delta\sigma_T = \int d\sigma(v >0) - \int d\sigma(v
<0)\,. \label{asymdef}
 \ee
We now focus on the two limiting cases for the helicity structure
of the odderon amplitude.

$\bullet$ Let SCHC hold for the $f_2$ meson production, i.e. the
dominant final state is with helicity 1 and $g_f^1\approx 1$. In
this case the main effect will be the FB asymmetry (\ref{xiasym}).

The phase space integration subject to the cut $t_{FB}$
(\ref{perplim}) gives
 \be
{d\Delta\sigma_{FB} \over dM^2} = 0.9{3\sqrt{5} \over 4}
\sqrt{\sigma_\rho \sigma_f} \cdot {\cal I}_{12}(M^2)\approx 1.5
\sqrt{\sigma_\rho \sigma_f} \cdot {\cal I}_{12}(M^2) \,.\label{FB}
 \ee

$\bullet$ Let the $f_2$ meson be produced in the SCHNC state with
helicity 2, i.e. $g_f^2\approx 1$. In this case the main effect
will be the transverse asymmetry (\ref{vasym}). The phase space
integration subject to the cut $t_T$ (\ref{perplim}) yields
 \be
{d\Delta\sigma_T \over dM^2}= 0.507\cdot{9\sqrt{5} \over 16}
\sqrt{\sigma_\rho \sigma_f} \cdot {\cal I}_{12}(M^2)\approx
0.64\sqrt{\sigma_\rho \sigma_f} \cdot {\cal I}_{12}(M^2)\,.
\label{perp}
 \ee

The charge symmetric background is a sum of the Pomeron and the
odderon cross sections.  Since the odderon amplitude is considered
to be very small, it contributes negligibly to the background even
far from the $\rho$ peak,  and the charge symmetric background can
be approximated by the $\rho$ contribution
 \be
{d\sigma_{bkgd} \over dM^2} =\sigma_\rho
|D_1(M^2)|^2\times\left\{\begin{array}{ccl} 0.9\;&\mbox{ for }
FB&\;\;(|t|>0.1B_\rho^{-1})\,,\\[2mm]
 0.367&\mbox{ for } T&\;\;(|t|>B_\rho^{-1}) \,.\end{array}\right.
     \label{bckg}
 \ee

Below we report $\Delta\sigma_{FB}$ and $\Delta\sigma_T$
integrated over the suitable $M^2$ region.

We present our results in the form of the statistical significance
of asymmetries, which is defined as the ratio of the signal to
statistical fluctuations of background:
 \be
 SS_a=\fr{N_S}{\sqrt{N_B}}=\fr{{\cal L}\Delta\sigma_a}{\sqrt{{\cal
 L}\sigma_B}}\qquad(a=FB\mbox{ or } T )\,, \label{SS}
 \ee
where ${\cal L}$ is the effective luminosity integral for the
$\gamma p$ collisions.  We see no reasons for the recording of
scattered electrons or protons.  Therefore, for the sake of
definiteness, we take the luminosity ${\cal L}=0.1$ pb$^{-1}$,
implying that various detection efficiencies are absorbed into
this quantity. (This ${\cal L}$ corresponds approximately to one
million detected events under the $\rho$ peak.)

Let us discuss first values of statistical significance for cross
sections averaged over small interval of $M^2\pm\Delta M^2$,
$SS(M^2)$ (note that $SS$ under interest does not coincide with
$\int SS(M^2)dM^2$). According to eqs.~(\ref{overlap}) and
(\ref{bckg}) for both FB and T asymmetries,
 \be
SS_a(M^2)\propto \fr{{\cal I}_{12}(M^2)}{|D_1(M^2)|}\equiv \fr{Im(
D^*_2D_1 e^{i\delta_{\Pom\od}})}{|D_1|} \le |D_2|\,.\label{estss}
 \ee
Therefore the largest values of this  $SS(M^2)$ are located
within the $f_2$ peak. It is illustrated by Fig.~2, where local
values of these $SS_a(M^2)$ are shown in arbitrary units. Hence,
\begin{figure}[!htb]
   \centering
   \epsfig{file=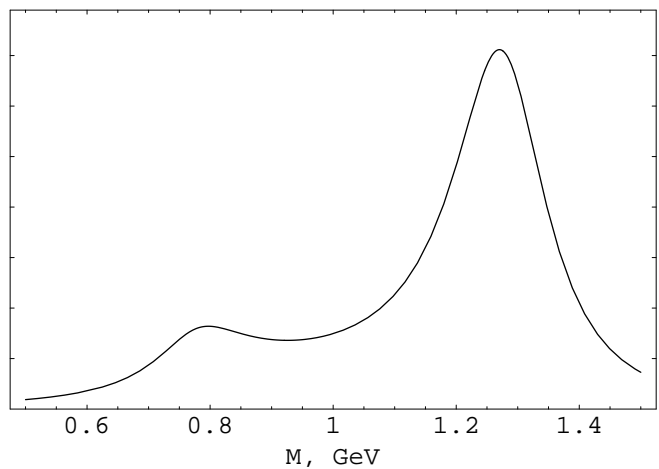,width=10cm}
      \caption{The local statistical significance
$SS_a(M^2)$ in arbitrary units.}
   \label{fig2}
\end{figure}
to obtain the best value of $SS$, we consider signals and
background integrated over the reasonable mass interval around the
$f_2$ peak. A natural choice\fn{The bump at $M\approx m_\rho$ is
not very useful, since the asymmetry in this region is also
affected by the interference with scalar resonances.} is
 \be
 M_f-\Gamma_f\;<M\;< M_f+\Gamma_f\,.\label{massint}
 \ee
Estimate (\ref{estss}) shows that the influence of nonresonant
background as well as tails of other resonances in the Pomeron
channel changes our estimates of $SS$ only weakly. Certainly, such
very estimate for interference with $S$--wave $\pi^+\pi^-$ final
states produced by odderon will show that the corresponding
signals are located near $f_0(600)$ and $f_0(980)$ peaks, and they
are negligible at $M>1100$ MeV.

With $\sigma_f = 1$ nb, integrating the asymmetries within
the region (\ref{massint}) yields
\be
\Delta\sigma_{FB} = 15.7\mbox{ nb}\,;\quad \Delta\sigma_{T} = 6.6\mbox{ nb}\,,
 \ee
which must be compared to the background cross section
 \be
\sigma_B=428\mbox{ nb for FB}\,,\quad \sigma_B= 174\mbox{ nb for T
}\,.\label{bckgcut}
 \ee
The statistical significance of the asymmetries for the luminosity
${\cal L}=0.1$ pb$^{-1}$ equals
\be
SS_{FB} \approx 7.5\,;\quad SS_T \approx 5.0\,.
 \ee
These number are still very promising, despite the fact that we
used very cautious estimates both for $f_2$ photoproduction cross
section and for the integrated luminosity. This offers certain
confidence that the odderon signal is indeed within the reach of
the current experiments even with very low value for the odderon
induced cross section and luminosity used here for the estimates.

\section{Discussion}

The charge asymmetry of dipions in diffractive photoproduction
$\gamma + p \to \pi^+\pi^-+p'$ emerges as a very attractive
signature for the odderon exchange. In the absence of competitive
mechanisms, an observation of such an asymmetry will be an
unambiguous discovery of the odderon.

As far as the discovery potential of the HERA experiments on $ep$
collisions is concerned, the main contribution to the dipion comes
from the quasireal photons. We see no reasons for either tagged
photons or going to deep inelastic regime with strongly suppressed
cross section. One must simply focus on the dipion final states
separated by a large rapidity gap from the scattered proton or its
low mass excitations.

With modest estimates for the $f_2$ production cross section
($\sigma(\gamma p \to f_2 p) \ge 1$ nb), the statistical
significance still turns out high ($SS\ge 5\div 8$) with very
cautious estimate for effective $\gamma p$ luminosity, ${\cal
L}_{\gamma p} = 0.1$ pb$^{-1}$. This suggests that the odderon
signal is definitely within the reach of the HERA experiments.

Certainly, the observation of charge asymmetry of pions in the
dipion mass region below $1.1$ GeV will be also unambiguous signal
of the odderon, and it can be even larger than that estimated above. 
However, a detailed calculation in this mass region
demands knowledge of specific models e.g. for the interaction of
odderon to different $f_0$ resonances, which cannot be developed
unambiguously now (for example, this coupling can be reduced
strongly if some of these resonances have large admixture of
gluonium). Therefore, in this paper we only suggest to look for
charge asymmetry for the discovery of the odderon. The subsequent detailed
analysis of the charge asymmetry at $M<1.1$ GeV (predominantly
transverse) can be used for the study of the coupling of different
$f_0$'s to the odderon. It can be the subject of separate paper(s).

The analysis of charge asymmetries proposed here is a very general
tool for extracting new information whenever the production
mechanism involves the both $C=+1$ and $C=-1$ exchanges (see
\cite{review} for other problems).

$\bullet$ {\bf Photon polarization dependence.} If the photons are
supposed to arise from electrons in $ep$ collider, then, in the
case of unpolarized initial electrons, the photons will be
linearly polarized in the electron scattering plane. This
modification changes neither the value nor the shape of the charge
asymmetry, and only introduces an overall factor that depends on
$\psi$.

For the longitudinally polarized electrons, the photons acquire
circular polarization. In this case the effective overlap function
acquires additional $\phi$ dependence, which leads to
non-universal helicity--sensitive azimuthal--charge asymmetry
\cite{Ter11}.

$\bullet$ {\bf A brief comment on the breaking of quark--hadron
duality}. Our optimism on the $\pi^+\pi^-$ production is based on
the observation that at the hadronic level the Pomeron--odderon
interference is strongly enhanced by the final-state interaction.
The study of the charge asymmetry in the $c\bar{c}$ final state
for the odderon discovery was proposed by \cite{brodsky}.
Certainly, the observation of this asymmetry requires observation
of both $c$ quarks (i.e. $D$ or $D^*$ mesons and pions). Near the
open charm threshold the essential part of these final states is
given by $DD$ or $DD^*$ or $D^*D^*$ states with well defined
effective mass, which allows one to discuss charge asymmetry in a
definite form. In this region, such final states come from decays
of a number of known C-odd and yet undiscovered  C--even
$c\bar{c}$ resonances (effect of FSI). The overlapping of these
resonances should enhance the effect essentially as compared to
that at the partonic level \cite{brodsky} at least near the open
charm threshold. This is the same FSI effect that was discussed
above for the dipions, and it exemplifies a general statement that
{\em in the charge asymmetry the quark--hadron duality can be
badly broken even for heavy quarks}.\\

We are thankful to S. Brodsky, A. Denner, H. Jung, S. Levonian, B.
Pire, V.G. Serbo, L. Szymanowski, O.V. Teryaev for useful
discussions. This paper is supported by grants RFBR 02-02-17884 \&
00-15-96691 and INTAS 00-00679 \& 00-00366.

\section*{ Appendix: Non-odderon contributions}

The two sources of the non-odderon $C$-even diffractive dipion
production are the $\omega$ and $\rho$ reggeon exchanges, and the
photon exchange --- the Primakoff effect. The estimates below show
that the odderon induced effect with $\sigma_f\ge 1$ nb with cuts
(\ref{perplim}) cannot be mimicked by these non-odderon
mechanisms.

$\bullet$  {\em The $\rho/\omega$ reggeon exchange contributions}
are estimated via the $\gamma p \to f_2 p$ at a fixed target
photon energy $\nu_{0} = 6$ GeV ($s_{0} = 12.1$ GeV$^2$) equals
$\sigma^{\rho/\omega}(\gamma p \to f_2 p) \approx 120$ nb (see
\cite{Shestakov}). Taking the $\rho/\omega$ reggeon trajectory
from \cite{PDG}, this cross section can be extrapolated to higher
energies as $ \sigma^{\rho/\omega}(s) \approx
\sigma^{\rho/\omega}(s_0) \cdot (s_0/s)^{0.9}$. At a typical HERA
energy $\sqrt{s} = 200$ GeV this yields
\be
\sigma^{\rho/\omega}(\gamma p \to f_2 p)_{HERA} \approx 0.15
\mbox{ nb}\,.
\label{rhoomega}
 \ee

$\bullet$ {\it For the Primakoff one-photon exchange  $f_2$
production} the cross section can be estimated reliably in the
equivalent photon approximation in terms of the two--photon width
of resonance $\Gamma_{\gamma\gamma}$  \cite{BGMS}:
 \be
d\sigma^{Pr}=\fr{8\pi\alpha \Gamma_{\gamma\gamma}(2J+1)}{M^3}\cdot
\fr{\vec{\Delta}^2 d\vec{\Delta}^2}
{(\vec{\Delta}^2+Q^2_m)^2}\;\quad\mbox{with}\quad
Q^2_m=\left(\fr{m_pM^2}{s}\right)^2\,.
 \ee
The integration over $\vec{\Delta}^2$ is effectively limited from
above by the proton form--factor at $\vec{\Delta}^2\sim m_\rho^2$.
It leads to a large logarithm $L=2\ln(m_\rho s/m_p M_f^2) \approx
15$ in the total cross section, $\sigma_{tot}^{Pr} \approx$ 8 nb.
This large cross section is concentrated strongly near forward
direction. If we impose the lower cut of $|\vec{\Delta}|^2=
B_\rho^{-1} \approx 0.1$ GeV$^2$, then the logarithmic enhancement
goes down more than one order in the magnitude, yielding the
Primakoff background cross cross section  $\lsim 0.5$ nb in the
observation region.

It is useful to note that Primakoff effect can produce $f_2$ meson
only in the states $\lambda_f=2,\,0$ (and it is experimentally
confirmed that the $\lambda_f=2$ dominates). Since SCHC holds for
$\rho$ meson production, it means, according to the above
analysis, that the main charge asymmetry from Primakoff--Pomeron
(P$\Pom$) interference will be transverse (T) while the P$\Pom$ FB
asymmetry is low, especially at small $|\Delta|$. Therefore, in
the study of FB asymmetry under interest one can use much lower
cut in $\vec\Delta$ (\ref{perplim}).

This estimate together with (\ref{rhoomega}) sets an approximate
lower limit of the odderon cross section, for which the observed
charge asymmetry must be regarded as an unambiguous evidence for
the odderon.

$\bullet$ As we see, in the region we are interested (\ref{perplim}), the
Primakoff contribution is expected to be smaller than the odderon
signal. However, at very low momentum transfer,
$|\vec{\Delta}|<50\div 100$ MeV, the Primakoff contribution will
dominate over the odderon one. Thus, the data from this region can
give us information about the phase of the forward $\gamma p\to
\rho p$ amplitude ("Pomeron phase") \cite{GINP}.

$\bullet$ Finally, we mention ref.~\cite{Kur} which discusses
charge asymmetry in electroproduction of dipions. Authors
consider the C--even dipion production only via Primakoff
mechanism and the C--odd dipion production only via the
bremsstrahlung mechanism. The latter is negligible compared to
the dominant Pomeron exchange completely overlooked in \cite{Kur}.

\end{document}